\documentclass[twocolumn,graphicx,pra]{revtex4}
\usepackage{amssymb}
\usepackage{epsfig}
\usepackage{bm}
\usepackage{amsmath}
\usepackage{graphicx}
\usepackage{epstopdf}

\begin{document}

\title{Quantum-orbit analysis of high-order harmonic generation by resonant
plasmon field enhancement}
\author{T. Shaaran$^1$, M. F. Ciappina$^1$ and M. Lewenstein$^{1,2}$}
\affiliation{$^1$ICFO-Institut de Ci\`{e}nces Fot\`{o}niques, 08860 Castelldefels,
Barcelona, Spain\\
$^2$ICREA-Instituci\'o Catalana de Recerca i Estudis Avan\c{c}ats, Lluis Companys,
08010 Barcelona, Spain}
\date{\today}

\begin{abstract}
We perform a detailed analysis of high-order harmonic generation (HHG) in atoms within the strong field
approximation (SFA) by considering spatially inhomogeneous monochromatic laser fields. We investigate
how the individual pairs of quantum orbits contribute to the harmonic spectra. We show
that in the case of inhomogeneous fields, the electron tunnels with two different canonical momenta. One
of these momenta leads to a higher cutoff and the other one develops a lower cutoff.
Furthermore, we demonstrate that the quantum orbits have a very different behavior in comparison to the homogeneous field.
We also conclude that in the case of the inhomogeneous fields, both odd and even harmonics are present
in the HHG spectra. Within our model, we show that the HHG cutoff extends far beyond the standard semiclassical cutoff
in spatially homogeneous fields. Our findings are in good agreement both with
quantum mechanical and classical models.
\end{abstract}

\maketitle

\section{INTRODUCTION}

\label{introduction}
In the context of the interaction of matter with
strong laser fields, high-order harmonics generation (HHG) process~\cite{McPherson1987,Huillier1991} has attracted considerable interest,
since it represents a viable route to the generation
 of coherent radiation in the ultraviolet (UV) to extreme
ultraviolet (XUV) spectral range with high repetition rate. Due to this high repetition rate, as well as high coherence degree and wavelength tunability,
HHG has found numerous applications in various areas of science, such as material
sciences, life sciences and lithography~\cite{Krausz2009}.

In addition, HHG has been employed for generating ultrashort pulses, and even single
attosecond pulses~\cite{Scrinzi2004}. This allows even more control over the atomic and molecular processes.
For instant, by superposing the XUV pulses on the laser field, one can resolve
dynamic processes in atoms or molecules with precision of a few attoseconds~\cite{Drescher2002}.
The physics behind the HHG can be understood by a
simple semi-classical three-step model~\cite{Corkum1993}. According to this
picture, an electron leaves the atom or molecule by tunneling through the
potential barrier, formed by the atomic potential and the laser electric field, to reach the continuum.
It subsequently propagates in the continuum and driven back by the laser filed toward its parent ion or
molecule. Finally, upon its return, it recombines with the core and leads to the
emission of energetic photons.

The threshold intensity for generating high-order harmonic in noble gases is
above $10^{13}$ W/cm$^{2}$, which is far beyond the output power of the current
femtosecond oscillators. Nowadays, chirped-pulse amplification (CPA) is used to
exceed the threshold intensity. CPA is a complex process and it requires
multi-pass amplifier cavities in tandem. On the other hand, as far as the
applications of the HHG are concerned, the optimization of HHG efficiency and
the extension of the cutoff to short-wavelength region are important aspects to consider~\cite{Niikura2005,Itatani2004}.
 The cutoff frequency of the generated harmonics can be extended either by reducing (increasing) the laser frequency (wavelength) or increasing the
peak field amplitude. There is, however, limitations in tuning these
parameters. In fact, reducing the laser frequency leads to a significant
drop in the harmonic yield due to the longer electron excursion times \cite{Tate2007,Colosimo2008},
while increasing the laser field intensity produces depletion of the ground state.

A recent demonstration has shown that the surface plasmon resonance could
provide a possible solution to overcome the problems cited above~\cite{Kim2008}. The
local field enhancement induced by a resonant plasmon within a metallic
nanostructure requires no extra cavities or laser pumping for amplification.
In this scheme, the local electric fields can be intensified by more than 20 dB \cite{Muhlschlegel2005,Schuck2005}, an amplification that exceed the threshold
laser intensity for HHG generation in noble gases. In addition, the pulse
repetition rate remains unchanged without adding any additional cavities or
extra pumping. Moreover, each nanostructure acts as a point-like source to
generate harmonics radiation, which through constructive interference can
provide even more focused coherent radiation. This gives a wide range of
possibilities to spatially rearrange nanostructures to shape or enhance HHG
spectral and even obtain a new physics. The locally enhanced field, has a
distinct spatial dependency, which gives an enormous extension to the HHG
cutoff~\cite{CiappinaNHPA2012}.

HHG based on plasmonics can be understood as follows~\cite{Kim2008}: the
external femtosecond low intensity pulse couples to the plasmon mode and
induces a collective oscillation of free charges within the localized
regions of the nanostructure. The free charges redistribute the electric
field around the nanostructure vicinity, in such, to form a spot of highly enhanced
electric field. The enhanced field, which largely depends on the geometrical
shape of the metallic nanostructure, exceeds the threshold intensity required
for HHG. As a result, by injection of noble gases into the spot of the
enhanced field, one can generate high order harmonics. In Ref.~\cite{Kim2008},
the output of the femtosecond oscillator, which was a pulse with 10 fs pulse
duration, 800 nm of wavelength and intensity of 10$^{11}$ W/cm$^{2}$, was
directly focused onto a bow-tie nanoantenna. As a result of the laser pulse
interaction with the nanostructure, the field intensity is enhancement by 2-4
order of magnitude, which is sufficient enough to produce XUV wavelengths
from the 7$^{th}$ (114 nm) to the 21$^{st}$ (38 nm) harmonics in Argon.

Hitherto, the theoretical approaches for studying strong field phenomena are
largely based on the assumption that the laser electric field is spatially
homogeneous in the region where the electron motions take place~\cite{Protopapas1997,Brabec2000}.
This assumption, however, does not hold for the
field enhanced by resonant plasmons. Indeed, the strong confinement of the
electrons in the plasmonic hot spots generates a spatially inhomogeneous
electric field, which strongly influences the subsequent motion of the electrons in the
strong field phenomena. As a result, new physics will emerge in the
interaction between matter and strong laser fields.

Since the first observation of HHG, different theoretical models, including solving
the Time Dependent Schr\"{o}dinger Equation (TDSE) and the Strong
Field Approximation (SFA), have been applied to describe this phenomenon (for
details see the review articles in~\cite{Saileres1999,LHuillier2008}). On
the other hand, the HHG by resonant plasmon field enhancement is a new topic
and it has been considerably less well studied since it is far more
difficult to measure and model. Up till now, two experiments have been performed to measure HHG based on this kind of field~\cite{Kim2008,Kimnew}, while only few theoretical approaches
have been developed~\cite{Hosakou,Yavuz2012,CiappinaNHPl2012}. 
One should note, however, that the interpretation of Ref~\cite{Kim2008} remains controversial~\cite{sivis,Kimreply,corkum_priv}.  

%we computed it by
%solving TDSE in one dimension~\cite{CiappinaNHPl2012}.

In our previous paper~\cite{CiappinaNHPA2012}, in which we employed SFA to investigate
the HHG by resonant plasmon field enhancement, the vector potential field was defined in an approximated way starting from the
inhomogeneous electric field. In the present paper, we improve our model even more by applying the same approximation directly
to the inhomogeneous electric field rather than the potential.
Subsequently, the vector potential is now derived from the electric field. In
addition, we use a SFA based on saddle point methods rather than a full numerical SFA approach to
obtain the HHG spectra. The saddle point methods lead to equations that can
be directly related to the classical equations of motion of an electron in a
laser field. As a result, they provide a space-time picture which gives us
additional physical insight. In this work, we
scrutinize the individual electron trajectories in comparison to their
classical counterparts and demonstrate their contributions to the HHG spectra.
In addition, since the imaginary part of the saddle point equations can be
related to the width of the potential barrier through which the electron
tunnels, we examine the ionization probability of the electron for each
trajectory.

This article is organized as follows. In Sec. II, we present the SFA transition amplitude for high-order harmonics,
starting from common expressions based on homogeneous electric fields (Sec. II.A) and,
subsequently,  by showing how we modify it for the case with non-homogeneous
fields (Sec. II.B). In Sec. III, we discuss the saddle-point
equations and analyze them in terms of quantum orbits in parallel to their
classical counterpart electron trajectories. In the next section, IV, we present
the HHG spectra based on the analysis given in Sec. III. Finally, in Sec. V,
we summarize the paper and state our main conclusions.

\section{THEORY}

\label{theory}

\subsection{Transition Amplitude}

\label{transitionamp}

Generally, there are two main assumptions behind the strong field
approximation (SFA), namely (i) the influence of the laser field is neglected when the
electrons are bound to their target atoms and (ii) the binding ionic potential is neglected
when the electrons are in the continuum. As a result, the free electrons in
the continuum are described by field-dressed plane waves, which are known as
Volkov states \cite{Gordon1926,Volkov1935}.

\subsubsection{Homogeneous fields}

\label{homogenoufield}

In the Lewenstein model \cite{Lewenstein1994}, a well established SFA-based method
to model HHG, it is assumed that the laser
electric field does not change with respect to the position in the region
where the electron motion takes place. In this spatially homogeneous field,
the SFA transition amplitude for HHG reads (in
atomic units)

\begin{eqnarray}
b_{\Omega } &=&i\int_{-\infty }^{\infty }\hspace*{-0.2cm}dt\hspace*{-0.1cm}%
\int_{-\infty }^{t}\hspace*{-0.3cm}dt^{\prime }\hspace*{-0.1cm}\int
d^{3}kd_{rec}^{\ast }(\widetilde{\mathbf{k}}(t))d_{ion}(\widetilde{\mathbf{k}%
}(t^{\prime }))  \notag \\
&&e^{-iS(\Omega ,\mathbf{k},t,t^{\prime })}+c.c.  \label{Mp}
\end{eqnarray}
with the action
\begin{equation}
S_{0}(\Omega ,\mathbf{k},t,t^{\prime })=\int_{t^{^{\prime }}}^{t}\hspace{%
-0.1cm}\frac{[\mathbf{k}+\mathbf{A}(\tau )]^{2}}{2}d\tau
+I_{p}(t-t^{^{\prime }})-\Omega t  \label{action}
\end{equation}
and the prefactors
\begin{equation}
d_{ion}(\widetilde{\mathbf{k}}(t^{\prime }))=\langle \mathbf{\tilde{k}}%
(t^{\prime })|H_{int}(t^{\prime })|\phi _{0}\rangle  \label{prefion}
\end{equation}
\begin{equation}
d_{rec}(\widetilde{\mathbf{k}}(t))=\langle \mathbf{\tilde{k}}%
(t)|O_{dip}.e_{x}|\phi _{0}\rangle .  \label{prefrec}
\end{equation}
Thereby, $k$, $I_{p}$, $\Omega $ , $\ H_{int}(t^{\prime })$, $O_{dip}$ and $%
e_{x}$ denote the drift momentum of the electron in the continuum, the
ionization potential of the of the field-free bound state $\left\vert \phi
_{0}\right\rangle $, the harmonic frequency, the interaction of the system
with the laser field, the dipole operator and the laser polarization vector,
respectively. The vector potential $\mathbf{A}(t)$ of the laser electric
field $E(t)$ is defined by
\begin{equation}
\mathbf{A}(t)=-\int_{-\infty }^{t}E(t^{\prime })dt^{\prime }.
\label{potentialfield}
\end{equation}
Physically, Eq. (\ref{Mp}) describes a process in which an electron, initially
in a bound state $|\phi _{0}\rangle $ with energy $I_{p}$, interacts with
the laser field by $H_{int}(t^{\prime })$ at the time $t^{\prime }$ and
tunnels into a Volkov state $|\mathbf{\tilde{k}}(t)\rangle $. Subsequently,
from time $t^{\prime }$ to $t$, it propagates in the continuum and is driven
back by laser field to its parent ion. At the time $t$, upon its return,
this electron recombines with the core and emits high-harmonic radiation of
frequency $\Omega $. The second term in Eq. (\ref{Mp}), which corresponds to
the continuum-continuum transitions, can be ignored, since it contributes
very insignificant to the transition amplitude of HHG (for details see \cite{BeckerHHG1997} ).
The main drawback of SFA is that it is not gauge invariant.
As a result, the matrix elements describing the ionization and
recombination, i.e. Eqs (\ref{prefion}) and (\ref{prefrec}), have different form
in the length and velocity gauges. This comes from the fact, that both the
interaction Hamiltonian $H_{int}(t^{\prime })$ and the Volkov wave function $%
|\widetilde{\mathbf{k}}(t)\rangle$ are not translationally invariant. The
interaction Hamiltonian is given by $H_{int}^{l}(t)=\mathbf{r}\cdot\mathbf{E}(t^{\prime })$ and $%
H_{int}^{v}(t)=[\mathbf{k}+\mathbf{A}(t^{\prime })]/2$ in the length and velocity gauges,
respectively. For the Volkov wave function,\ $\widetilde{\mathbf{k}}(t)=%
\mathbf{k}+\mathbf{A}(t)$ in the length gauge and $\widetilde{\mathbf{k}}(t)=\mathbf{k%
}$ in the velocity gauge.
In this paper, we work in the length gauge and we assume that the electric
field is linearly polarized along $x$-axis. Furthermore, we consider an hydrogenic
$1s$ state for the field-free bound state $\left\vert \phi _{0}\right\rangle$.

As a result, the Eqs. (\ref{prefion}) and (\ref{prefrec}) yield
\begin{equation}
d_{ion}(\widetilde{\mathbf{k}}(t^{\prime }))\propto \frac{\widetilde{\mathbf{%
k}}(t^{\prime })_{x}}{(\widetilde{\mathbf{k}}(t^{\prime })^{2}+\alpha
^{2})^{3}}E(t^{\prime })  \label{newprefion}
\end{equation}

\begin{equation}
d_{rec}(\widetilde{\mathbf{k}}(t))\propto \frac{\widetilde{\mathbf{k}}(t)_{x}%
}{(\widetilde{\mathbf{k}}(t)^{2}+\alpha ^{2})^{3}}  \label{newprefrec}
\end{equation}

\subsubsection{Nonhomogeneous fields}

\label{nonhomogeneousfield}

We will now consider a case in which that assumption made in Sec. \ref%
{homogenoufield} is not any more valid and the laser field has a spatially
inhomogeneous character, when the HHG process takes place. Before discussing
the nonhomogeneous case, we examine how the action of the SFA is connected to
classical electron trajectories for the homogeneous field. The laser potential
$V_{L}$ due to the laser field $E(t)$ is defined as
\begin{equation}
V_{L}=xE(t),  \label{potfieldhm}
\end{equation}
and the Newton equation of motion for an electron in this field is given by
\begin{equation}
\ddot{x}(t)=-\nabla _{x}V_{L}  \label{Newtonho}
\end{equation}%
In here, the force is equal to (minus) the laser electric field, i.e. $\ddot{x}(t)=-E(t)$.  In the SFA, the action is defined in terms of the vector potential field $A(t)$ given by (\ref{potentialfield}), which is the counter
part of the velocity $\dot{x}(t)$.

For the inhomogeneous case the electric field now has the form $E(t,x)$ and the laser
potential is $V_{L}=xE(t,x)$. Thus, the Newton equation of motions become
\begin{equation}
\ddot{x}(t)=-x\nabla _{x}E(t,x)-E(t,x).  \label{Newton}
\end{equation}
From Eq. (\ref{Newton}), it is clear that $\ddot{x}(t)\neq -E(t,x)$. Therefore,
the potential field $A(t)$ of the SFA action should correspond to the
integration of $\ddot{x}(t)$ with respect to $t$.

If the spatial dependence of the laser electric field is perturbative and linear with
respect to position, then the field can be approximated as
\begin{equation}
E(t,x)\simeq E(t)(1+\epsilon x),  \label{nhmfield}
\end{equation}
where $\epsilon \ll 1$ is a parameter that characterize the strength of the inhomogeneity.

Indeed, the above approximation corresponds to the first term of the actual
field of a plasmonic nanostructure with spherical shape \cite{Klingnano2011}. By substituting (\ref{nhmfield}) into (\ref{Newton}), we have
\begin{equation}
\ddot{x}(t)=-E(t)(1+2\epsilon x(t)).  \label{Newtonh1}
\end{equation}
This is the effective laser electric field that the electron feels
along the trajectory $x(t)$, which describes its motion in the continuum. We will call it the electron trajectory effective electric field.

Classically, the electron trajectory can be found by solving Eq (\ref{Newtonh1}). In here, we solve it by applying the Picard iteration \cite{EdwardG1999} method and restrict ourselves to the first order (for more details see \cite{CiappinaNHPA2012} ). Based on the condition that the
electron starts its movement at the origin with zero velocity, i.e. $x(0)=0$
and $\dot{x}(0)=v(0)=0$ , we obtain
\begin{equation}
x(t)=\beta (t)-\beta (t_{0})-A(t_{0})(t-t_{0}).  \label{position}
\end{equation}
with $\beta (t)=\int_{0}^{t}dt^{\prime }A(t^{\prime })$. In
addition we assume that at time $t_{0}$ the potential field is zero; thus
\begin{equation}
x(t)=\int^{t}dt^{\prime }A(t^{\prime }).  \label{newposition}
\end{equation}

By using Eqs. (\ref{potentialfield}), (\ref{newposition}) and (\ref{Newtonh1}),
the effective vector potential along the electron trajectory $A_{tr}(t)$ reads
\begin{equation}
A_{tr}(t)=A(t)+2\epsilon A_{c}(t),  \label{newpotentialfield}
\end{equation}
where
\begin{equation}
A_{c}(t)=\int^{t}dt^{\prime \prime }A(t^{\prime \prime })-\int^{t}dt^{\prime
\prime }A^{2}(t^{\prime \prime }).
\end{equation}
The next step is to modify the general expression of the transition amplitude (\ref{Mp}) for the HHG, in
order to include the above defined inhomogeneous field. Consequently, we have to replace the
electric field and vector potential by Eqs. (\ref{nhmfield}) and (\ref{newpotentialfield}), respectively. As a result, the modified action yields
\begin{eqnarray}
S(\Omega ,\mathbf{k},t,t^{\prime }) &=&S_{0}(\Omega ,\mathbf{k},t,t^{\prime
})+2\epsilon \int_{t^{^{\prime }}}^{t}\hspace{-0.1cm}A_{c}(\tau )[\mathbf{k}+%
\mathbf{A}(\tau )]d\tau  \notag \\
&&+2\epsilon ^{2}\int_{t^{^{\prime }}}^{t}\hspace{-0.1cm}A_{c}^{2}(\tau
)d\tau  \label{newaction}
\end{eqnarray}
where $S_{0}(\Omega ,\mathbf{k},t,t^{\prime })$ is defined in Eq. (\ref{action}).

\subsection{Saddle-point equations}

\label{saddleEqs}

The transition amplitude (\ref{Mp}) can be computed either numerically or
using the saddle-point method~\cite{Bleinstein1986,Salieres2001}. In here,
we employ the latter procedure since the solutions of the saddle point
equations are directly related to the classical trajectories. Thus, it allows us
to investigate the quantum orbits in comparison to the classical
trajectories as well as demonstrating their contributions to the cutoff and
yield of the HHG. This method requires obtaining the saddle points where the
action (\ref{newaction}) is stationary, i.e. for which
$\partial _{t}S(\Omega ,\mathbf{k},t,t^{\prime })=\partial _{t^{\prime }}S(\Omega ,\mathbf{k},t,t^{\prime })=\partial _{k}S(\Omega ,\mathbf{k},t,t^{\prime })=0$. In this
paper, we use a specified steepest descent method called uniform
approximation to take care of those saddles points which are not well separated
(for a detailed discussion see Ref. \cite{Faria2002} ). The stationary
conditions upon $t,t^{\prime }$ and $k$ lead to the saddle-point equations
\begin{equation}
\left[ \mathbf{k}+\mathbf{A}(t^{\prime })\right] ^{2}-2\epsilon \lambda (t^{\prime })=-2I_{p}  \label{saddle1}
\end{equation}
\begin{equation}
\int_{t^{\prime }}^{t}d\tau \lbrack \mathbf{k}+\mathbf{A}(\tau )]+2\epsilon
\eta \mathbf{(\tau )}=0  \label{saddle2}
\end{equation}
and
\begin{equation}
\Omega =\frac{\left[ \mathbf{k}+\mathbf{A}(t)\right] ^{2}}{2}%
+I_{p}+2\epsilon \lambda (t)  \label{saddle3}
\end{equation}
with
\begin{equation}
\lambda (t)=A_{c}(t)[\mathbf{k}+\mathbf{A}(t)]+\epsilon
A_{c}^{2}(t),  \label{saddle4}
\end{equation}
and
\begin{equation}
\eta(\tau )=\int_{t^{^{\prime }}}^{t}\hspace{-0.1cm}A_{c}(\tau
)d\tau .  \label{saddle5}
\end{equation}%
Eq. (\ref{saddle1}) expresses the conservation law of energy for the electron
tunnel ionized at the time $t^{\prime }$. Eq. (\ref{saddle2}) guarantees that
the electron returns to its parent ion as well as constraining the
intermediate momentum of the electron. Finally, (Eq. \ref{saddle3}) gives the
energy conservation of the electron at the time $t$, when upon its return
recombines with the core and releases a high frequency photon $\Omega $.

The terms $\lambda(t^{\prime})$, $\eta(t)$ and
$\lambda(t)$ in the Eqs. (\ref{saddle1}), (\ref{saddle2}), and (\ref{saddle3}), respectively, emerge from the nonhomogeneous character of the laser field
and they vanish for the homogeneous case, i.e when $\epsilon =0$. For the
homogeneous case, the solutions of the saddle equations are generally complex since Eq. (\ref{saddle1}) admits no real solutions,
unless $I_{p}\rightarrow 0$. This is a consequence of the fact that tunneling has no classical
counterpart. For the inhomogeneous case, however, it is not very upfront to
constrain the limit, in which the solutions of (\ref{saddle1}) are real.
Nevertheless, in here, $\epsilon$ is a very small parameter and the electron will most
likely reach the continuum with tunnel ionization. Thus, the solutions of
these saddle point equations are still expected to be complex. In addition, the maximum
kinetic energy that the electron gains in the continuum is not any more $3.17U_{p}$, where $U_{p}=E_{0}^{2}/(4\omega ^{2})$ is the ponderomotive
energy. In fact, it depends on the nonhomogeneous character of the field, i.e. of $\epsilon$ and $A_{c}(t)$. For positive $A_{c}$, the electron gains
energy, depending on the value of the $\epsilon $, larger than $3.17U_{p}$,
while for the negative $A_{c}$, it would be below the conventional value.

We now examine the drift momentum $\mathbf{k}$ of the electron at the
time of the tunneling Eq. (\ref{saddle1}) . For that, we consider the limit $I_{p}\rightarrow 0$, where the electron reaches the continuum with zero
kinematical momentum. As a result, Eq. (\ref{saddle1}) yields to
\begin{equation}
\mathbf{k}=-A(t^{\prime })+\epsilon A_{c}(t^{\prime })(1\mp \sqrt{3})
\label{driftmomentum}
\end{equation}
Unlike the homogeneous case, where $\mathbf{k}=-A(t^{\prime })$, in here,
$\mathbf{k}$ has two different solutions with one exceeding and the other
lowering the homogeneous drift momenta. The strength of the inhomogeneity $\epsilon$ and the shape of the $A_{c}(t^{\prime })$ are the responsible of this
deviation.

\begin{figure}[h]
\begin{center}
\includegraphics[width=8cm]{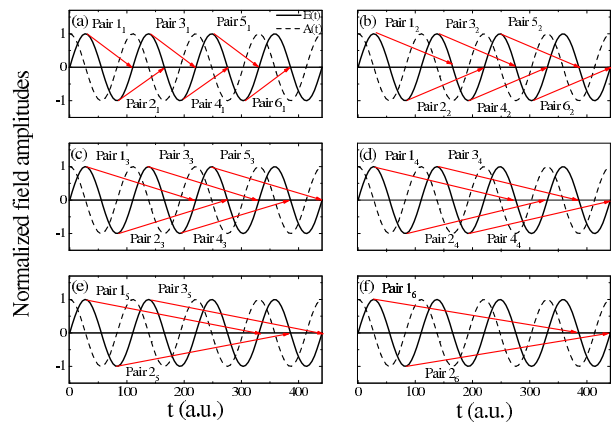}
\end{center}
\caption{Schematic representation of the laser electric field $\mathbf{E}(t)$ and
the corresponding potential field $\mathbf{A}(t)$ for a monochromatic
field defined by $\mathbf{E}(t)=E_0 \sin(\omega t) e_{x}$, with 4
cycles. The arrow indicates the classical times around which the electrons
leave to the continuum and return at the field crossing, approximately. The
pairs of the orbits are indicated by the labels $Pair_{n}$, where $n$ range from
$1$ to $8$. Panels, a,b,c,d,e and f represent the first, second, third,
fourth, fifth and sixth shortest pairs, respectively. The fields are
normalized to $\mathbf{A}(t)/A_{0}$ and $\mathbf{E}(t)/E_{0}$ , where $A_{0}$
and $E_{0}$ are the field amplitudes.}
\label{pulse}
\end{figure}

\section{RESULTS}

\label{results}

\subsection{Quantum orbits}

\label{gmorbits}

In this section, we investigate the role of individual trajectories to the HHG
cutoff for the nonhomogeneous case by performing a quantum-orbit analysis of
the problem. The concept of the quantum-orbits is based on the fact that the
solutions of the saddle-point equations can be related to the classical
trajectories of the electron and, in addition, to obtain information on quantum
aspects such as tunneling and interference. To get a better insight into the nonhomogeneous case, we employ a monochromatic field with $E(t)=E_{0}\sin
(\omega t)e_{x}$, where $e_{x}$  is the polarization vector along the
$x$-axis. By using the relationship defined in Eq. (\ref{potentialfield}) and
applying some trigonometric identities, the laser effective electric field (\ref{Newtonh1})
and effective potential field (\ref{newpotentialfield}) along the electron trajectory read
\begin{equation}
E_{tr}(t)=E_{0}\sin (\omega t)(1+2\epsilon \sin (\omega t)/\omega ^{2}),
\label{monochelectricfield}
\end{equation}
\begin{equation}
A_{tr}(t)=A_{0}\cos (\omega t)+2\epsilon A_{c}(t),
\label{monochpotentialfield}
\end{equation}
respectively, where $A_{0}=E_{0}/\omega$ and
\begin{equation}
A_{c}(t)=A_{0}^{2}\sin (\omega t)/4\omega -A_{0}^{2}t/2.
\label{monochcorecpotentialfield}
\end{equation}
In terms of the pondermotive energy of the homogeneous field, the drift momentum of Eq. (\ref{driftmomentum}) yields
\begin{equation}
\mathbf{k}=-2\sqrt{U_{p}}\cos (\omega t^{\prime })+\epsilon (\frac{U_{p}}{%
\omega }\sin (\omega t^{\prime })-2U_{p}t^{\prime })(1\mp \sqrt{3})
\label{newdriftmomentum}
\end{equation}

\begin{figure}[h]
\begin{center}
\includegraphics[width=6cm]{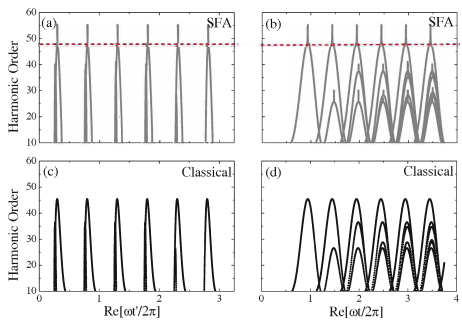}
\end{center}
\caption{(Color online) Dependence of harmonic order on the release time $t^{\prime }$ and the recombination time $t$
of the electron for all given pairs in Fig. (\ref{pulse}) and for the homogeneous field ($\epsilon=0$). We
consider an hydrogen atom, for which the ground-state energy is $I_{p}=0.5$ a.u.,
in a linearly polarized, monochromatic field of frequency $\omega=0.057$ a.u. and intensity $I=3\times10^{14} \mathrm{W/cm^{2}}$. Panels (a) and (b)
give the harmonic order as a function of the ionization and recombination times of the SFA model, respectively,
while panels (c) and (d) depict the the harmonic order in terms of ionization and recombination times of
the classical calculations, respectively. The red dashed lines correspond to the harmonic cutoff.}
\label{HHG1}
\end{figure}

Based on the above equations, we solve the saddle point equations defined in Eqs. (\ref{saddle1})- (\ref{saddle3}) in terms of the ionization $t'$ and recombination $t$ times. For more close analysis, we restrict ourselves just to the solutions of the first 4 cycles of our defined monochromatic field, as shown in Fig. \ref{pulse}. Classically, it is most probable that electron ionizes at the electric field maxima and returns to its parents ion at the electric field crossings. In Fig.\ref{pulse},  panels (a), (b), (c) and (d) depict the cases when the electron leaves at the field maxima and returns to the core at time about $\pi$, $2 \pi$, $3 \pi$, and  $4 \pi$ later, respectively.

\begin{figure}[h]
\begin{center}
\includegraphics[width=7.5cm]{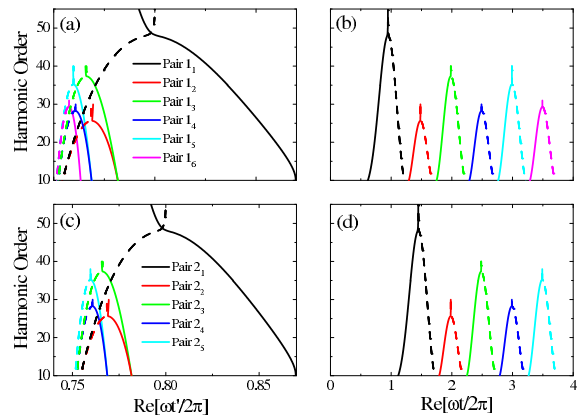}
\end{center}
\caption{(Color online) SFA harmonic order as a function of the real part of the release time
 $t^{\prime }$ and the recombination time $t$ of the electron for the
same parameters as in Fig. \ref{HHG1}, but for pairs $1_{n}$ and $2_{m}$ (where $n=1-6$ and $m =1-5$ ) of Fig. \ref{pulse}.
Panels (a) (from right to left $1_{1}\rightarrow 1_{6}$) and (b) (from left to right $1_{1}\rightarrow 1_{6}$) give the ionization and recombination times of pairs $1_{n}$, respectively. Panels (c) (from right to left $2_{1}\rightarrow 2_{5}$) and (d) (from left to right $2_{1}\rightarrow 2_{5}$) depict the ionization and recombination times of pairs $2_{m}$, respectively. The dashed and solid
lines correspond to the long and the short orbits, respectively.}
\label{HHG2}
\end{figure}

In Fig. \ref{HHG1}, we plot the the harmonic order as function of the real parts of the ionization $t'$ and recombination $t$ times for the case with $\epsilon=0$ (panels a and b, respectively). In this figure, in comparison to the SFA model, we also present the classical solutions of $t'$ and $t$ (panels c and d, respectively). From Fig. \ref{HHG1}, it is clear that the SFA resemble the classical calculations. Apart from that, both calculations show that the ionization and recombination times corresponding to each cycle are identical.

\begin{figure}[h]
\begin{center}
\includegraphics[width=7.5cm]{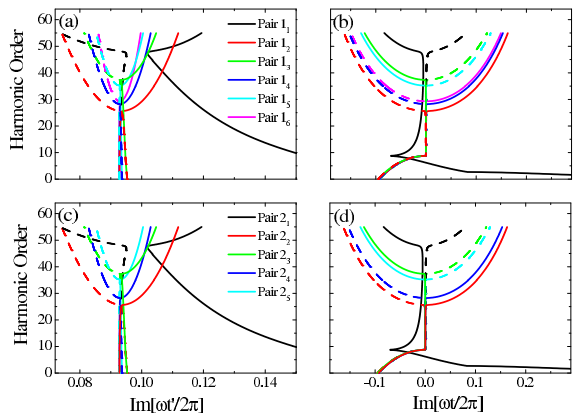}
\end{center}
\caption{(Color online) SFA harmonic order as a function of the imaginary part of the release time
 $t^{\prime}$ and the recombination time $t$ of the electron for the
same parameters as in Fig. \ref{HHG1}, but for pairs $1_{n}$ and $2_{m}$
(where $n=1-6$ and $m =1-5$ ) of Fig. \ref{pulse}. Panels (a) and (b) give the
ionization and recombination times of pairs $1_{n}$,
respectively, and panels (c) and (d) depict the ionization and
recombination times of pairs $2_{m}$, respectively. The dashed and solid
lines correspond to the long and the short orbits.}
\label{HHG3}
\end{figure}

\begin{figure}[h]
\begin{center}
\includegraphics[width=7.5cm]{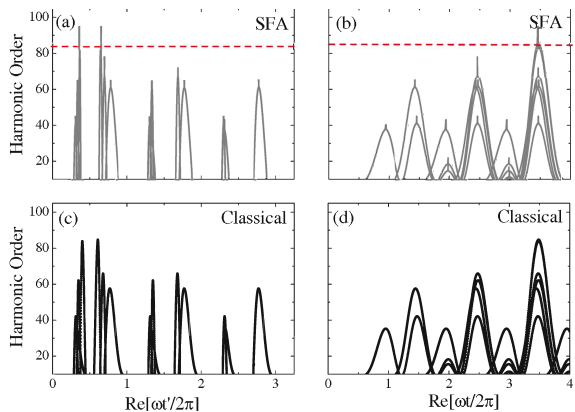}
\end{center}
\caption{(Color online) Dependence of harmonic order on the release time $t^{\prime }$ and the recombination time $t$ of the electron for all given
pairs in Fig. \ref{pulse}, for a nonhomogeneous field with $\epsilon=0.003$.
We consider hydrogen atoms for which the ground-state energy is $I_{p} = 0.5$
a.u. in a linearly polarized, monochromatic field of frequency $\protect%
\omega=0.057$ a.u. and intensity $I=3\times10^{14}\mathrm{W/cm^2}$. Panels (a) and
(b) give the ionization and recombination times of SFA model, respectively,
while panels (c) and (d) depict the ionization and recombination times of
the classical calculations, respectively. The red dashed lines correspond
to the harmonic cutoff.}
\label{HHG4}
\end{figure}

In Fig. \ref{HHG2}, we isolate those solutions and plot the harmonic order in terms of the real parts of ionization and recombination times by considering the case when the electron leaves the atom at the times around $\pi/2$ and returns to the core at the times $n\pi$ ($n=2,3,4,5$) (panels (a) and (b) for $t'$ and $t$, respectively) and the case when the electron leaves the atom at the times around $\pi$ and returns to the core at $n\pi$ ($n=3,4,5$) (panels (c) and (d) for $t'$ and $t$, respectively). For a given harmonic, there is always a
shorter (solid line) and a longer (dashed line) travel time for the electron in the continuum, corresponding to the long and short trajectories of the pair. Such pairs of orbits coalesce at the maximally allowed harmonic energies, i.e at the cutoff.  In here, the shortest orbit, i.e. the one that the electron leaves at the field maxima and returns at around time $\pi$ later, have the largest cutoff, at harmonic $47\omega$. It means that these pairs of orbits lead the cutoff of the HHG spectra, while the others pairs produce harmonics with lower energies.

In Fig. \ref{HHG3}, we present the imaginary parts of ionization and recombination times of the pairs shown in Fig. \ref{HHG2}. Panels (a) and (b) depict the $t'$ and $t$ for $n \pi$ ($n=2,3,4,5$), respectively. Panels (c) and (d) demonstrate the $t'$ and $t$ for $n \pi$ ($n=3,4,5$), respectively. For the recombination times, $Im[t]$ essentially vanishes between the harmonic order for which the real parts $Re[t]$ coalesce. Physically, this means that, in this region, the recombination is classically allowed. Beyond this region, $Im[t]$ increases abruptly, which indicates that the classically forbidden region has been reached. On the other hand, the imaginary part $Im[t']$ of the start time of the electron is always non-vanishing. This is due to the fact that the tunneling has no classical counterpart. These results show that both the imaginary and real parts of the tunneling and recombination times for pairs $1_{n}$ ( where $n=1-6$ ) and $2_{m}$ ($m =1-5$ ) given in Fig. \ref{pulse} are similar.

\begin{figure}[h]
\begin{center}
\includegraphics[width=7.5cm]{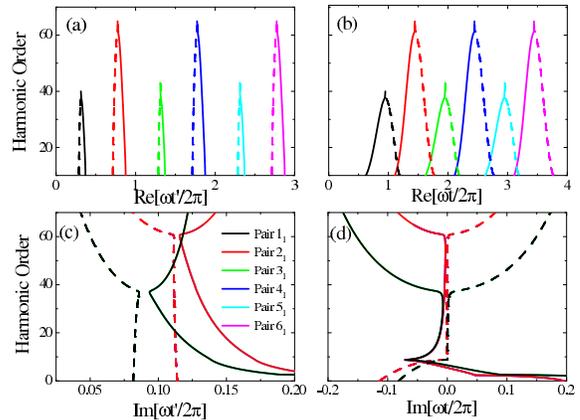}
\end{center}
\caption{(Color online) Dependence of harmonic order on the release time $t^{\prime }$ and the recombination time $t$ of the electron for the
same parameters as in Fig. \ref{HHG4}, but for pairs $n_{1}$ (where $n=1-6$) of Fig. \ref{pulse}.
Panels (a) and (c) (from left to right $1_{1}\rightarrow 6_{1}$) depict the real part of the release and
recombination times, respectively, while panels (c) and (d) show the imaginary part of the ionization and
recombination times of pairs $n_{1}$, respectively. The dashed and solid
lines correspond to the long and the short orbits.}
\label{HHG6}
\end{figure}

Now we move to the nonhomogeneous case and consider $\epsilon=0.003$. In Fig. \ref{HHG4}, we demonstrate the harmonic order as function of the real parts of the ionization $t'$ and recombination $t$ times for such a case. Like above, the SFA (Panels (a) and (b) for $t'$ and $t$, respectively) resembles the classical calculations (Panels (c) and (d) for $t'$ and $t$, respectively). The general harmonic cutoff is extended in comparison to the homogeneous case, but the trajectories do not follow the same symmetry as shown in Fig. \ref{HHG1}.

To closely examine the orbits, we plot the harmonic order of the shortest pairs, i.e pairs $n \pi$ (n=1 to 5), in terms of ionization and recombination times (Fig. \ref{HHG6}). Panels (a) and (c) represent the real and the imaginary parts of $t'$, respectively and, panels (b) and (d) represent the real and the imaginary parts of $t$, respectively. Unlike the homogeneous case, these pairs do not lead to the same cutoff.  For the pairs corresponding to the electron leaving at the field maxima (Fig. \ref{pulse}(a)), the cutoff is at around harmonic $38\omega$, while for the pairs corresponding to the electron leaving at the field minima the cutoff is at around harmonic $60\omega$. These results come from the fact that, for a given harmonic, the electron may tunnel with two possible momenta given by Eq. (\ref{newdriftmomentum}). It appears that the electron has larger momenta if it tunnels from minima of the field and smaller momenta if it tunnels from maxima of the field.

\begin{figure}[h]
\begin{center}
\includegraphics[width=7.5cm]{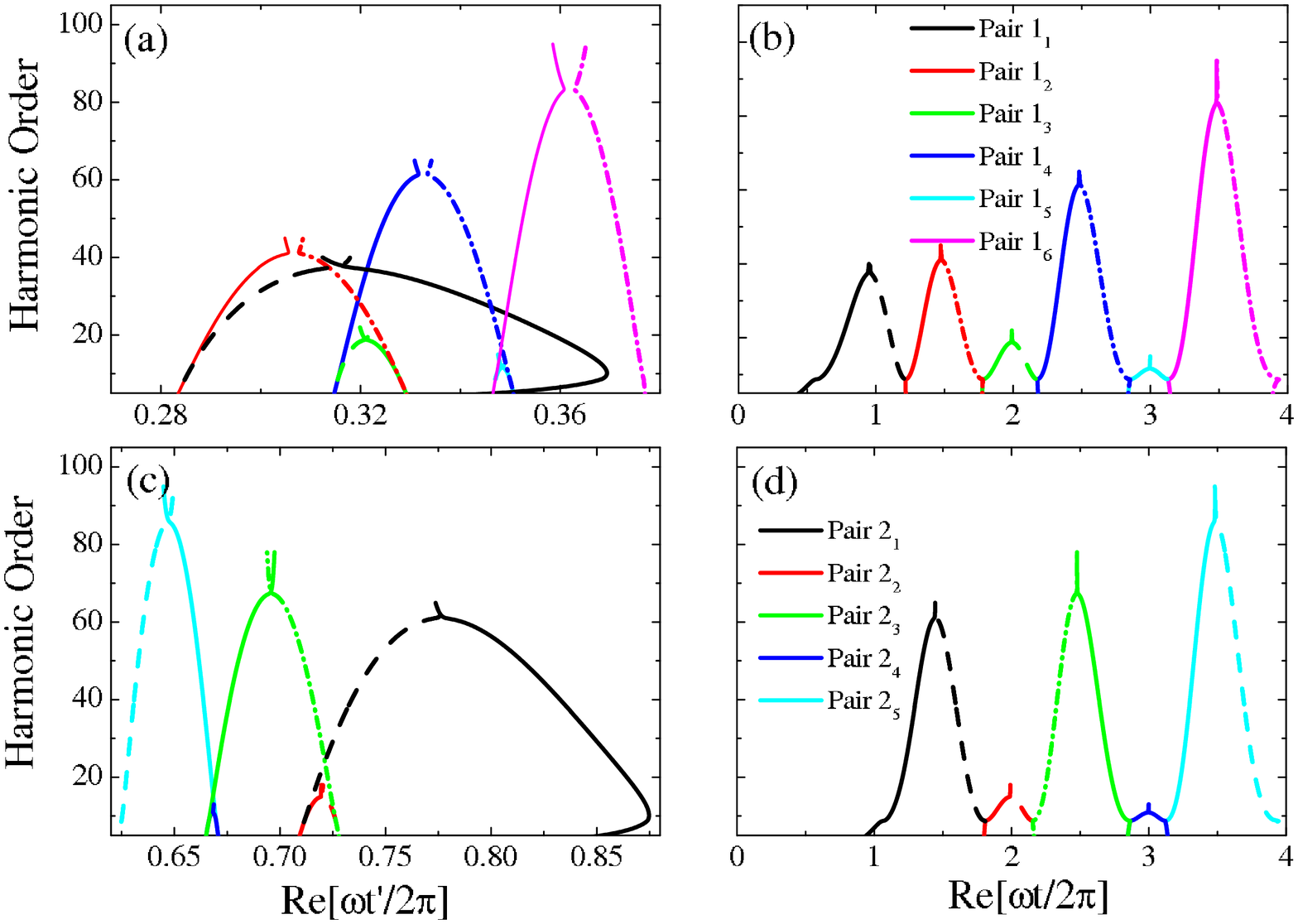}
\end{center}
\caption{(Color online) SFA harmonic order as a function of the real part of the release time
 $t^{\prime }$ and the recombination time $t$ of the electron for the
same parameters as in Fig. \ref{HHG4}, but for pairs $1_{n}$ and $2_{m}$ (where $n=1-6$ and $m =1-5$ ) of Fig. \ref{pulse}.
Panels (a) and (b) (from left to right $1_{1}\rightarrow 1_{6}$) give the ionization and recombination times of pairs $1_{n}$,
respectively, and panels (c) and (d) (from left to right $2_{1}\rightarrow 2_{5}$) depict the ionization and
recombination times of pairs $2_{m}$, respectively. The dashed and solid
lines correspond to the long and the short orbits,  while the pairs with dot dashed lines do not have the well-known shorts and long pairs.}
\label{HHG5}
\end{figure}

\begin{figure}[h]
\begin{center}
\includegraphics[width=6.5cm]{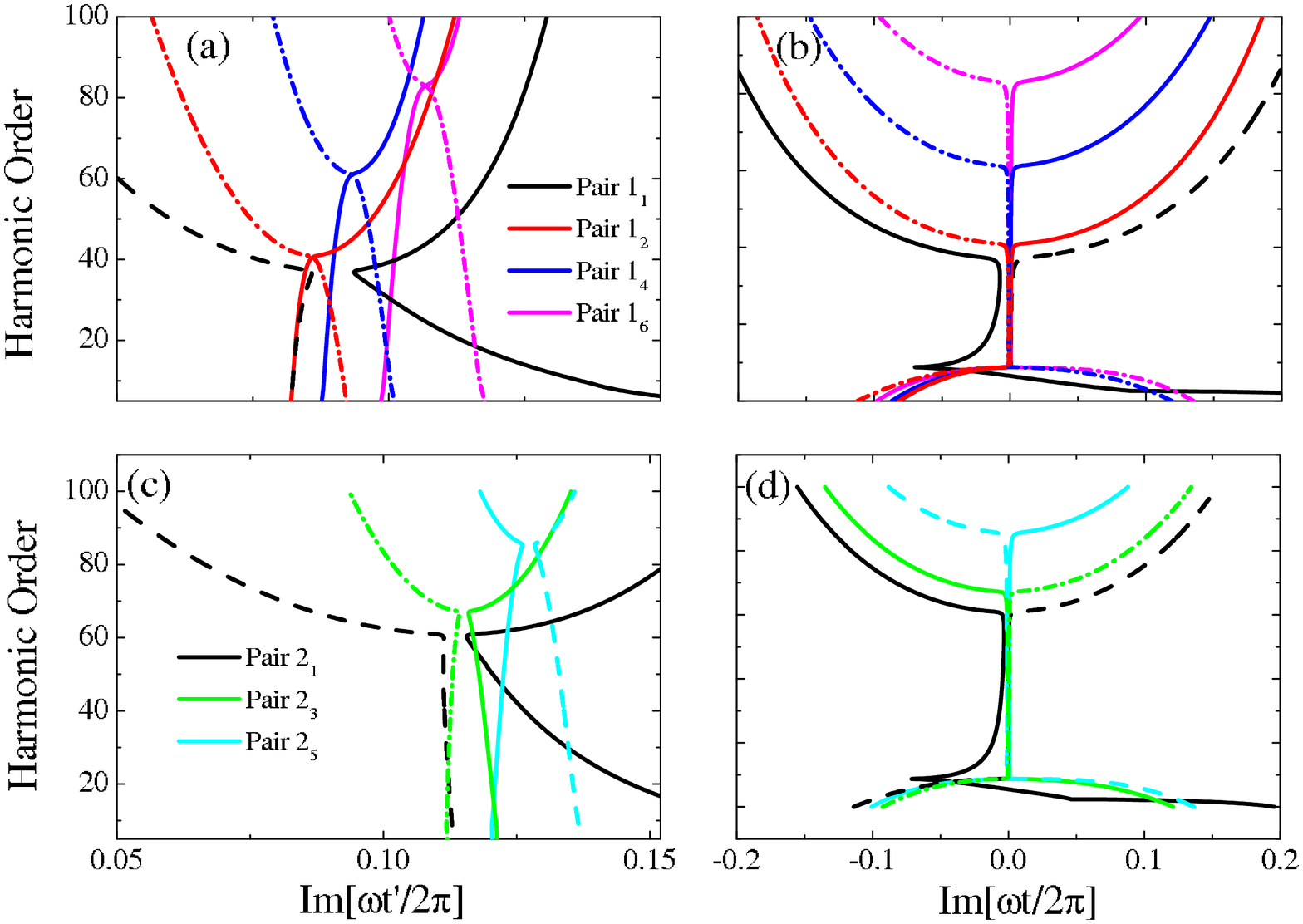}
\end{center}
\caption{(Color online) SFA harmonic order as a function of the imaginary part of the release
 time $t^{\prime }$ and the recombination time $t$ of the electron for the
same parameters as in Fig. ~\ref{HHG4}, but for pairs $1_{n}$ and $2_{m}$ (where $n=1,2,4,6$ and $m =1,3,5$) of Fig. \ref{pulse}.
Panels (a) and (b) give the ionization and recombination times of pairs $1_{n}$,
respectively, and panels (c) and (d) depict the ionization and
recombination times of pairs $2_{m}$, respectively. The dashed and solid
lines correspond to the long and the short orbits, while the pairs with dot dashed lines do not have the well-known shorts and long pairs.}
\label{HHG8}
\end{figure}

Figures \ref{HHG5} and \ref{HHG8} demonstrate the dependence of harmonic order on the real and imaginary parts of ionization and recombination times, respectively, by considering the case when the electron leaves the atom at the times around $\pi/2$ and returns to the core at the times $n \pi$ ($n=2,3,4,5$) (panels (a) and (b) for $t'$ and $t$, respectively) and the case when the electron leaves the atom at the times around $\pi$ and returns to the core at $n \pi$ ($n=3,4,5$) (panels (c)and (d) for $t'$ and $t$, respectively). 

For the dominant pairs the cutoff become larger as we move from shorter pairs to the longer pairs. For instance, pair $1_{1}$, which is associated to the electron which leave at the times around $\pi/2$ and recombined at the times $2 \pi$, has cutoff at harmonic $38\omega$, while the cutoff of pair $1_{6}$, which is associated to the electron which leave at the times around $\pi/2$ and recombined at the times $5 \pi$, is at harmonic $80\omega$. In contrast to the homogeneous case, in here, the pairs which associated with the electron which recombined after few cycle later from the time of its ionization lead to the larger cutoff.  Furthermore, for some of the pairs (shown with dot dashed lines) like pair $1_{6}$ and pair $2_{3}$ the conventional concept of the short orbit, in which the electron leaves a bit later and returns a bit earlier, and the long orbit, in which the electron leaves a bit earlier and returns a bit later, does not have any meaning. In fact, for these pairs, if the electron leaves a bit early then return a bit early and if it leaves a bit later then returns a bit later.

\begin{figure}[h]
\begin{center}
\includegraphics[width=6.5cm]{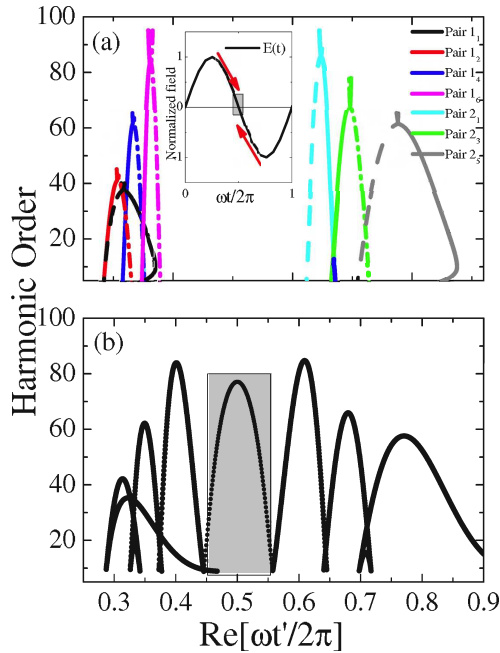}
\end{center}
\caption{(Color online) Harmonic order as a function of the release time
 $t^{\prime}$ of the electron for the
same parameters given in Fig. \ref{HHG4}, but for pairs $1_{n}$ and $2_{m}$ (where $n=1,2,4,6$ and $m =1,3,5$ ) of Fig. \ref{pulse}.
Panel (a) gives the real part of the release times based on SFA, while panel (b) depicts the classical ionization times of these pairs.
The dashed and solid lines correspond to the long and the short orbits, while the pairs with dot dashed lines do not have the well-known short and long pairs.}
\label{HHG7}
\end{figure}

In addition, the ionization time of the electron has even more interesting behavior as shown in Fig. \ref{HHG7}. For pairs $1_{n}$ and pairs $2_{m}$ (where $n=1-6$ and $m=1-5$ ), the ionization time $t'$ moves toward $\pi$ as $n$ and $m$ become larger, until both solutions collapsed on each other as it demonstrated in Fig. \ref{HHG8}. In this figure, panel (a) depict the SFA calculation while panel (b) represent the classical calculations. For the SFA model, the ionization time as function of $n$ and $m$ moves more slowly toward $\pi$ in comparison to the classical calculations. In the classical calculations, the collapse at $t'=\pi$ is associated to the pairs $1_{7}$ and $2_{6}$. In the SFA model, however, this collapse it is not exactly centered at around $t'=\pi$, instead it manifests itself by collapsing to its previous pair. On the other hand, both SFA and classical models give the same cutoff. It means the SFA will give a reliable HHG spectra while its yields will be affected quantitatively, which would not be a problem since SFA has the same limitation even for the homogeneous fields.

\begin{figure}[h]
\begin{center}
\includegraphics[width=7.5cm]{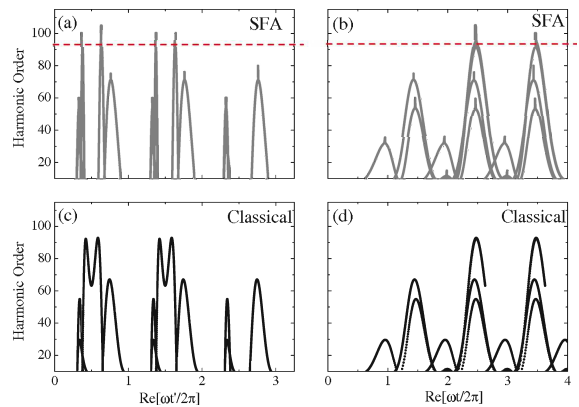}
\end{center}
\caption{(Color online) Dependence of harmonic order on the release time $t^{\prime }$ and the recombination time $t$ of the electron for all given
pairs in Fig. \ref{pulse}, for a nonhomogeneous field with $\epsilon=0.005$.
We consider hydrogen atoms for which the ground-state energy is $I_{p} = 0.5$
a.u. in a linearly polarized, monochromatic field of frequency $\omega=0.057$ a.u. and intensity $I=3\times10^{14}\mathrm{W/cm^2}$. Panels (a) and
(b) give the ionization and recombination times of SFA model, respectively,
while panels (c) and (d) depict the ionization and recombination times of
the classical calculations, respectively. The red dashed lines corresponds
to the harmonic cutoff.}
\label{HHG9}
\end{figure}

\begin{figure}[ht]
\begin{center}
\includegraphics[width=7.5cm]{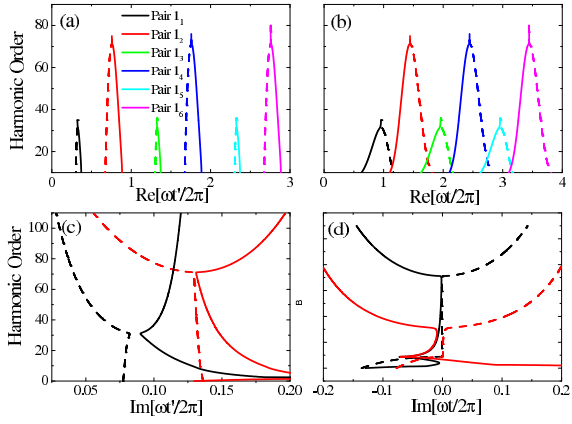}
\end{center}
\caption{(Color online) Dependence of harmonic order on the release time $t^{\prime }$ and the recombination time $t$ of the electron for the
same parameters as in Fig. \ref{HHG9}, but for pairs $1_{n}$ (where $n=1-6$) of Fig. \ref{pulse}.
Panels (a) and (b) (from left to right $1_{1}\rightarrow 1_{6}$) depict the real part of the release and
recombination times, respectively, while panels (c) and (d) show the imaginary part of the ionization and
recombination times of pairs $1_{n}$, respectively. The dashed and solid
lines correspond to the long and the short orbits, respectively.}
\label{HHG10}
\end{figure}

In Fig. \ref{HHG9}, we demonstrate the harmonic order as function of the real parts of the ionization $t'$ and recombination $t$ times for $\epsilon=0.005$. Panels (a) and (c) depict the $t'$ and panels (b) and (d) show $t$, for SFA and classical models, respectively. The SFA calculations are in good agreement with the classical model. In here, the general cutoff extended to the larger harmonic. In fact for the shortest pairs, the cutoff is at harmonic $70\omega$ and for the longest allowed pairs the cutoff extends to harmonic $92\omega$. In comparison to the case with $\epsilon=0.003$, for the shortest pairs ($1_{n}$ with $n=6$), the cutoff from maxima and the minima of the field shift towards lower (harmonic $32\omega$) and higher harmonic (harmonic $70\omega$) as demonstrated in Fig. \ref{HHG10}.

\begin{figure}[h]
\begin{center}
\includegraphics[width=7.5cm]{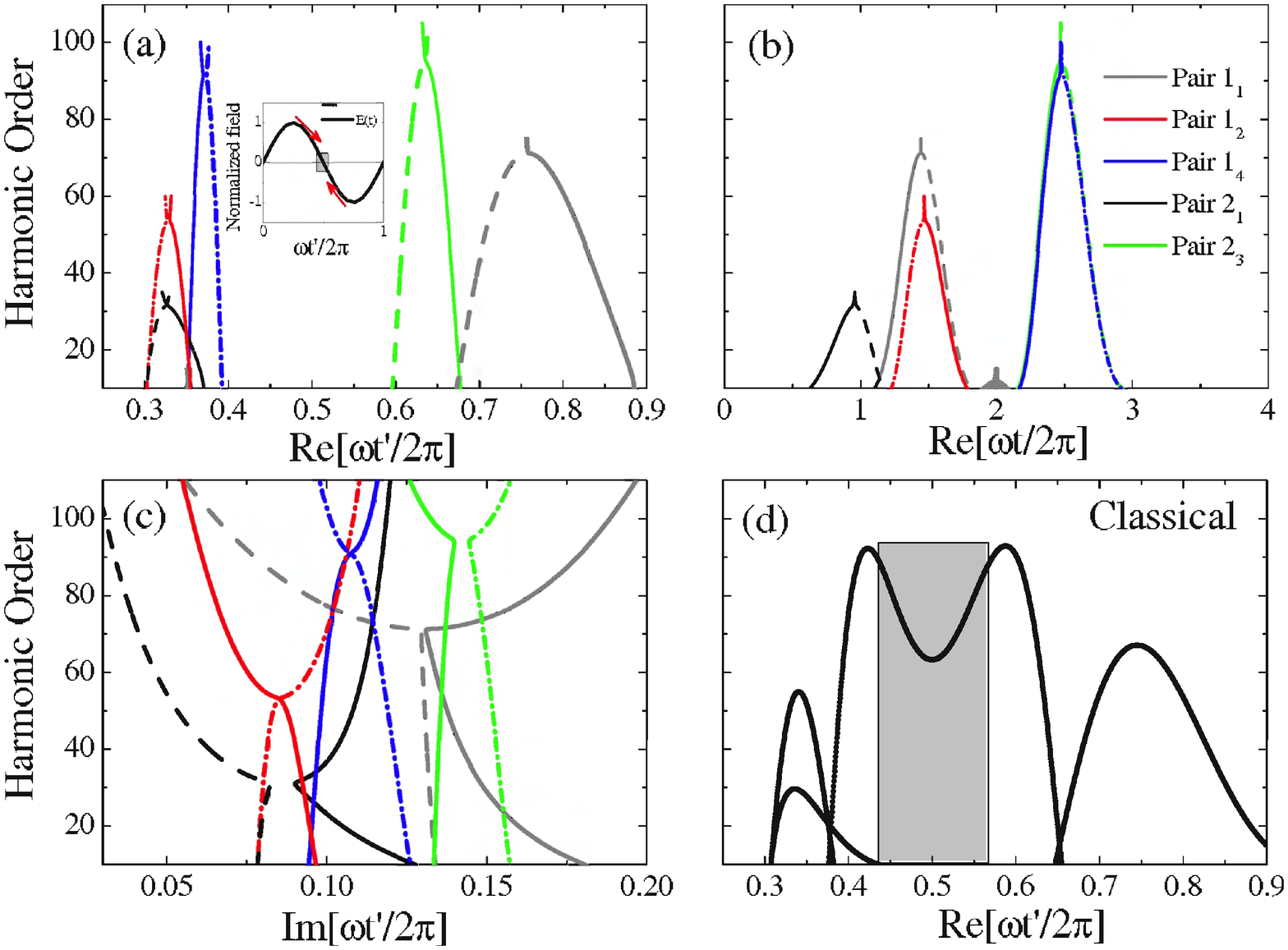}
\end{center}
\caption{(Color online) Harmonic order as a function of the release time
 $t^{\prime}$ and the recombination time $t$ of the electron for the
same parameters as in Fig. \ref{HHG4}, but for pairs $1_{1}, 1_{2}, 1_{4}, 2_{1}$ and $2_{3}$ of Fig. \ref{pulse}.
Panels (a) and (b) give the real part of the release and recombination times of pairs,
respectively, while panel (c) show their imaginary parts. Panel (d) depict the classical ionization times of these pairs. The dashed and solid
lines correspond to the long and the short orbits,  while the pairs with dot dashed lines do not have the well-known short and long pairs.}
\label{HHG11}
\end{figure}

Furthermore, the collapse of the pairs from maximum and minimum of a given cycle happens even faster as shown in Fig. \ref{HHG11}. As we discussed above for $\epsilon=0.003$, the collapse occurs at Pairs $1_{7}$ and $2_{6}$, while in here it is at Pairs $1_{4}$ and $2_{3}$. It means that, as field becomes more inhomogeneous, i.e  larger values of $\epsilon$, the electron does not return to the core if it follows the longer trajectories.

We verified, without showing it here, that for larger values of $\epsilon$  (like $\epsilon=0.02$ ) even the longer trajectory of the shortest pairs does not lead to return of the electron to the core. Despite the fact that our model does not accommodate such large values of $\epsilon$, it suggests that for strong inhomogeneous field just the shortest trajectories will lead to recombination process. In fact, our TDSE calculations~\cite{CiappinaNHPl2012}, which based on the actual nonhomogeneous field generated in the confined region of bow-tie nanostructures, show that the shortest trajectories rather than longest are contributing to the HHG spectra.

\subsection{Spectra}
\label{spectra}

\begin{figure}[h]
\begin{center}
\includegraphics[width=7.5cm]{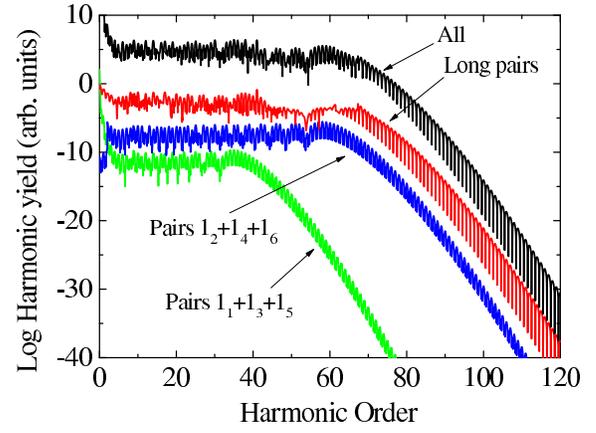}
\end{center}
\caption{(Color online) High-order harmonic spectra for hydrogen atoms ($I_{p}=0.5$ a.u.) and interacting with a monochromatic field of frequency $\omega=0.057$ a.u. and intensity $I=3\times10^{14} \mathrm{W/cm^{2}}$ for the case with $\epsilon=0.003$. Yellow and blue colored spectra show the contributions from the shortest pairs with smaller cutoff, i.e. pairs $1_{1}, 1_{3}$ and $1_{5}$, and the shortest pairs with largest cutoff, i.e. pairs $1_{2}, 1_{4}$ and $1_{6}$, while the red color demonstrates the contributions from all other orbits. Black colored spectra shows the total contributions from all pairs given in Fig. \ref{pulse}. For clarity, all the HHG spectra are scaled.}
\label{spectra1}
\end{figure}

In this section, we compute HHG spectra with Eq. (\ref{Mp}) and using the saddle point method developed in Section II.B. Fig. \ref{spectra1} presents HHG spectra for the case with $\epsilon=0.003$. The spectra with yellow and blue colors represent the contributions from the shortest pairs with smaller cutoff, i.e. pairs $1_{1}, 1_{3}$ and $1_{5}$, and the shortest pairs with largest cutoff, i.e. pairs $1_{2}, 1_{4}$ and $1_{6}$, while the red color demonstrates the contributions from all other orbits given in Fig. \ref{pulse}. For all three cases, the HHG cutoffs are in good agreement with the trajectories represented the previous section. The latter case, which corresponds to the longest pairs, leads to the largest cutoff. This is expected since for these pairs the electron has more time to accelerate in the field and return to the core with higher energy. On the other hand, for these pairs, the wave packet spreads too much in the continuum. Therefore, they lead to harmonic with lower yields. For the former case, which correspond to the shortest pairs, the harmonic yields will be large for both the cases, i.e. the one has largest and smallest cutoffs. The spectra with black color shows the total contributions, i.e from all the pairs given in Fig. \ref{pulse}.

\begin{figure}[h]
\begin{center}
\includegraphics[width=7.5cm]{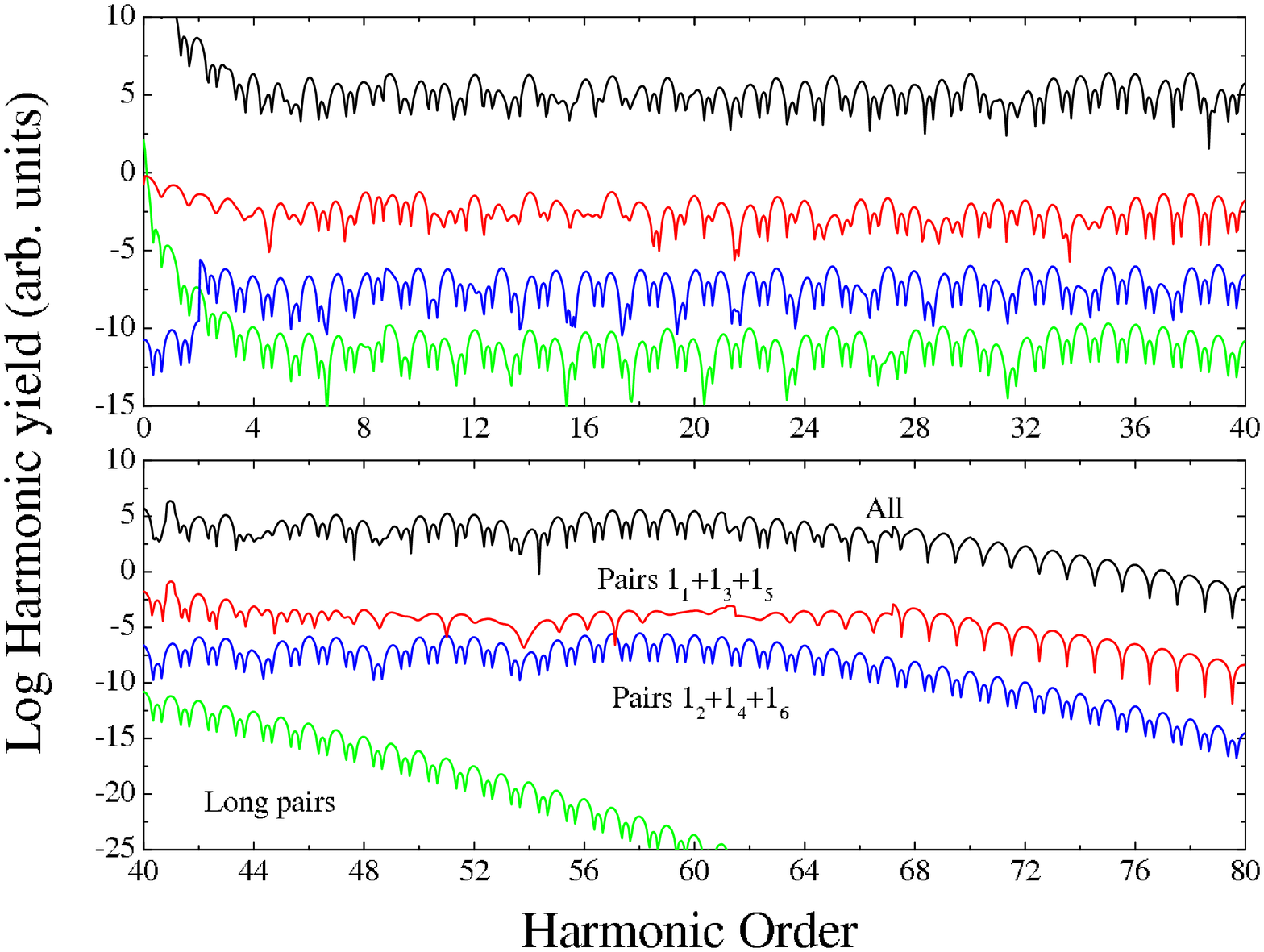}
\end{center}
\caption{(Color online) Zooming into the HHG spectra given in Fig. ~\ref{spectra1}}.
\label{spectra2}
\end{figure}

We now zoom into the spectra of Fig. \ref{spectra1} to examine more closely the harmonic generated from these sets of pairs (Fig. \ref{spectra2}). Up to harmonic $40\omega$, the shortest pairs with lower cutoff ($1_{1},1_{3}$ and $1_{5}$) give the shape of the total spectra and from harmonic $40\omega$ to $64\omega$ the shortest pairs with higher cutoff ($1_{2},1_{4}$ and $1_{6}$) dominate the shape of the total spectra. For higher harmonic the rest of the pairs lead the shape of the total spectra.

Furthermore, it seems that both odd and even harmonics are present in the total spectra. At the lower regime of the spectra, one sets of harmonics are a bit more dominant than the others while at higher regime both odd and even harmonic have the same weight.

\begin{figure}[h]
\begin{center}
\includegraphics[width=7.5cm]{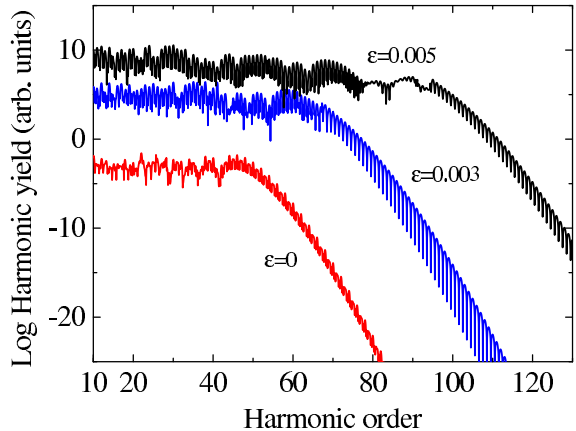}
\end{center}
\caption{(Color online) High-order harmonic spectra for hydrogen atoms ($I_{p}=0.5$ a.u.) and interacting with a monochromatic field of frequency $\omega=0.057$ a.u. and intensity $I=3\times10^{14}$ W/cm$^{2}$. Red, blue and black colors depict the cases with $\epsilon=0$, $\epsilon=0.003$ and $\epsilon=0.005$, respectively. All the HHG spectra are scaled for clarity.}
\label{spectra3}
\end{figure}

In Fig. \ref{spectra3} we compute the HHG spectra for homogeneous field (red colored spectra) and nonhomogeneous fields with $\epsilon=0.003$ (blue colored spectra) and $\epsilon=0.005$ (black colored spectra), respectively. For all three cases, the HHG cutoff is in good agreement with the trajectories analysis represented in the previous section and with the full 3D numerical calculations of Ref~\cite{Yavuz2012}. For the case with $\epsilon=0$, the cutoff is at around harmonic $45\omega$ and for $\epsilon=0.003$ the cutoff is at around harmonic $73\omega$. The field with $\epsilon=0.005$ leads to the largest cutoff at around harmonic $93\omega$. For the latter case, there is less interference in the region corresponding the harmonics $90\omega$ to $100\omega$. This behavior is expected since there are just two trajectories which contribute to such cutoff as shown in Fig. \ref{HHG9}.

\section{CONCLUSIONS}

\label{conclusion}

In this work, we show how the quantum orbits manifest themselves in spatially inhomogeneous fields. We show that in nonhomogeneous fields, the electron tunnels with two different canonical momenta: one leads to a higher cutoff and the other to a lower one. Furthermore, we demonstrate that for an electron tunneling at the field maxima the tunneling time becomes larger when it recombines at a later time, while for the electron tunneling at the field minima the tunnel time become smaller as it recombines at the later time. In fact, as we go from the shorter pairs to the longer ones, the trajectories from the minimum and maximum of the cycle moves towards each other until they collapse on each other at the field crossing. The nonhomogeneity character of the field determines this collapse. For larger nonhomogeneity strength the electron only returns to the core if it follows the shortest pairs of trajectories. In addition, for some of the trajectories we can not define the conventional pair of long and short orbits.  The former orbit corresponds to the case when electron leaves a bit earlier and returns a bit later and the latter case gives the trajectories in which an electron leaves a bit later and returns a bit earlier to the ionic core. Indeed, in here, if the electron leaves earlier it will then returns earlier and if it leaves later then it will return later. We also demonstrate that in the case of linear nonhomogeneous fields, both odd and even harmonics are present in the HHG spectra. At the lower harmonics one is a bit more dominant than the others while at higher harmonic order they both have equal weight. Within our model, we show that the HHG cutoff extends to the larger harmonics as a function of the inhomogeneity strength.

\section*{Acknowledgments}

We acknowledge the financial support of the MINCIN project FIS2008-00784 TOQATA (M. F. C. and M.L.); ERC Advanced
Grant QUAGATUA, Alexander von Humboldt Foundation and Hamburg Theory Prize
(M. L.). We thank Samuel Markson for useful comments and suggestions.

\end{document}